\documentclass[doublecol]{epl2}

\begin{document}

\title{Relaxation-to-creep transition of domain-wall motion in a two-dimensional random-field Ising model with an ac driving field}

\author{N.J. Zhou\inst{1,2}, B. Zheng\inst{1,2}\footnote{corresponding author; email: zheng@zimp.zju.edu.cn} and D.P. Landau\inst{1,3}}

\institute{ \inst{1} Zhejiang University, Zhejiang Institute of Modern Physics, Hangzhou 310027, PRC\\
 \inst{2}  Asia Pacific Center for Theoretical Physics, POSTECH, Pohang 790-784, Korea \\
\inst{3} Center for Simulational Physics, The University of Georgia,Athens, GA 30602, U.S.A.
}

\shorttitle {Relaxation-to-creep transition of domain-wall motion with an ac driving field}
\shortauthor {N.J. Zhou, B. Zheng and D.P. Landau}


\abstract {Monte Carlo simulations of a two-dimensional,random-field Ising model with an ac driving field are used to study the relaxation-to-creep transition of domain-wall motion at low temperatures. The resultant complex 
susceptibility $\chi = \chi' -i\chi''$ exhibits features in agreement with the experiments of ultrathin ferromagnetic and ferroelectric films: The semicircle and straight line in the $\chi'$-$\chi''$ plot are Cole-Cole signatures 
of relaxation and creep states, respectively. The exponent $\beta$ describing the creep motion is measured, and an intermediate state between the relaxation and creep states is identified.}

\maketitle

Domain-wall dynamics has become an important topic of study in magnetic devices, nanomaterials, thin films and semiconductors \cite{yam07,lem98,tan08,kim09}. In particular, quenched randomness in ferroic materials, e.g. ultrathin 
ferromagnetic and ferroelectric films, fundamentally affects the response to an external field \cite{bra05,met07,kle07,son08,dou08,im09,jo09}. From a purely theoretical point of view, the topic is also essential for understanding 
non-equilibrium dynamics in disordered media. For a dc (direct current) driving field, $H$, the domain-wall motion exhibits a depinning transition at zero temperature. The depinning field, $H_p$, separates the regimes of static 
pinning ($H < H_p$) and friction-limited viscous slide ($H> H_p$). At low, but non-zero, temperatures, the sharp depinning transition is softened and a thermally activated creep state appears 
\cite{now98,nat01,kol05,met07,yam07,bus08,zho09,jo09}.

Although the stationary states for a dc driving field are now understood, theoretical results for the dynamic states in an ac(oscillating) driving field, $H = H_0\exp(i2\pi f t)$ is limited. At low temperatures, four dynamic states 
(relaxation, creep, sliding and switching) may be classified in the so-called Cole-Cole diagrams \cite{che02,bra05,kle07,jez08}. Segmental domain-wall {\it relaxation} without net wall motion occurs at high frequencies and weak 
driving fields \cite{sny01,kle05,bra05,kle07,mel09}. For low frequencies and strong fields, the domain-wall dynamics turns into periodic {\it switching} between differently poled states \cite{rob08}. Between them, two different 
dynamic states are complemented \cite{gla03,kle05}. {\it Sliding} of a domain wall is driven by an external field with $H_0>H_p$, while {\it creep} is the {\it thermally} activated motion for weaker fields, $H_0<H_p$. 
Experimentally, all four dynamic states are observed in ferromagnetic and ferroelectric materials \cite{che02,kle07,kle07a}.

Theory suggested that a dynamic transition might separate the relaxation and creep states \cite{nat01}, and very recent, experimental evidence was found in ultrathin ferromagnetic and ferroelectric films, as well as in liquid 
crystals \cite{bra05,kle07,jez08}. \revision {Theoretically}, a Debye law and a power law may describe the complex susceptibility in the relaxation and creep states respectively \cite{sny01,fed04,kle07,nat90}.

Current theoretical approaches to domain-wall dynamics in ultrathin ferromagnetic and ferroelectric films are typically based on the Edwards-Wilkinson equation with quenched disorder (QEW) \cite{fed04,gla03,kle05,bra05}. The QEW 
equation is phenomenological, containing little microscopic information, and describes neither the relaxation state nor the relaxation-to-creep transition \cite{pet04}. To understand the domain-wall motion at a microscopic level, 
we need lattice models based on microscopic structures and interactions. The random-field Ising model with a driving field (DRFIM) is a candidate \cite{now98,rot01,col06,kle06b,cer08}, at least to capture certain robust features of 
domain-wall motion, although it is only a minimalistic model and does not \revision {include} all interactions in real materials. Very recently, the depinning transition of the DRFIM model was examined with the short-time dynamic 
approach, and the critical exponents were accurately determined \cite{zho09,zho10}. The domain-wall roughening process at the order-disorder phase transition has also been seen \cite{zho08,he09}.

In this Letter, our goal is to simulate the relaxation state and relaxation-to-creep transition of the domain-wall motion at low temperatures using the two-dimensional DRFIM model. Numerical results will be compared with those 
experiments of disordered ferromagnetic and ferroelectric films.

The two-dimensional DRFIM model is defined by the Hamiltonian
\begin{equation}
\revision {\mathcal{H} = - J\sum_{<ij>}S_iS_j -      \sum_i[h_i + H(t)]S_i},\label{equ10}
\end{equation}
where $S_i = \pm 1$ is a spin at site $i$ of a square lattice, $h_i$ is the quenched random field uniformly distributed within an interval $[-\Delta,\Delta]$, and $H(t)$ is a {\it homogeneous} driving field. For our study, 
\revision {we typically take the coupling constant $J=1$ and the disorder parameter $\Delta = 1.5$} \cite{zho09}, and perform simulations at low temperatures, $T$, with a periodic field $H(t) = H_0 \exp(i2\pi f t)$. When $H_0$ is 
sufficiently weak, the field strength dependence is negligible, so we \revision {usually fix $H_0 = 0.01$. However, extra simulations are also performed with other values of the parameters $J$, $\Delta$ and $H_0$ to examine their 
dynamic effects.} The range of frequencies is $10^{-4} < f < 10$ Hz. A total of $20 000$ runs were made so that statistical error bars in the results are usually smaller than the symbols used in the figures to follow. A total cpu 
time of 6 processor-years was used for the simulations.

The initial state is semi-ordered with a domain wall in the $y$ direction. Antiperiodic and periodic boundary conditions are used in $x$ and $y$ directions respectively. To eliminate the pinning effect irrelevant for disorder, we 
rotate the square lattice such that the initial domain wall is in the $(11)$ direction \cite{now98,zho09}. With this initial state, we update randomly selected spins with the heat-bath algorithm. Denoting $L_x$ and $L_y$ as the 
lattice sizes in $x$ and $y$ directions, a Monte Carlo time step is defined by $L_x \times L_y$ single-spin flips. As time evolves, the domain wall moves and roughens, while the bulk remains almost unchanged since the temperature 
is low. The domain interface width grows slowly, so lattice size $L_x = 12$ and $L_y = 512$ is used. Data for $L_x = 24$ confirm that the finite-width effect is negligible.

To describe the domain interface, we measure the magnetization from the height function
\begin{equation}
M(t) = \left \langle \sum^{L_x}_{x = 1}S_{xy}(t)\right \rangle.\label{equ20}
\end{equation}
Here $S_{xy}$ denotes a spin at site $(x,y)$, and $<\ldots>$ represents both the statistic average and average in the $y$ direction. After the stationary magnetization hysteresis loop is obtained at $t > t_{0}$, the complex 
susceptibility $\chi = \chi' -i \chi''$ is calculated by \cite{pet04, fed04}
\begin{equation}
\chi(f,T) = \frac{1}{PH_0}\int_0^{P}dtM(t)e^{-i2\pi f t},    \label{equ40}
\end{equation}
where $P= 1 /f$ is the period of the ac driving field. In our simulations, $t_{0}$ is about $20$ periods, and another $20$ periods are performed for calculating $\chi$. To extract the creep motion induced by the temperature, we 
introduce a pure susceptibility
\begin{equation}
D\chi(f,T)=\chi (f,T)-\chi (f,T=0).\label{equ50}
\end{equation}
For the weak driving field, $M(t)$ oscillates by
\begin{equation}   M(t) = A \cos(2\pi f t - \delta) + \overline M.\label{equ60}
\end{equation}
Here $A$ is the amplitude, $\delta$ is the phase shift and $\overline M$ is the average magnetization. Substituting Eq.~(\ref{equ60}) into Eq.~(\ref{equ40}), we find
\begin{equation}
 A(f,T) = 2H_0\sqrt{\chi'^2+\chi''^2},   \quad \quad   \tan [\delta (f,T)] = \frac{\chi''}{\chi'}.\label{equ70}
\end{equation}

\revision {Theoretically}, the Debye law derived from non-interacting two-level systems may characterize the {\it relaxation} state at $T=0$ or at high frequencies and $T>0$ \cite{sny01,bra05,kle07,nat90},
\begin{equation}
\chi'=\frac{\chi_{\infty}(1/f)^2}{(1/f)^2+\tau^2},   \quad    \quad   \chi''=\frac{\chi_{\infty}\tau (1/f)}{(1/f)^2+\tau^2},\label{equ80}
\end{equation}
where $\tau$ is a relaxation time and $\chi_{\infty}=\chi'(1/f=\infty)$. It gives rise to a semicircle in the $\chi'$-$\chi''$ (Cole-Cole) plot.  With Eqs.~(\ref{equ70}) and (\ref{equ80}), one may obtain
\begin{equation}
A \propto \frac{(1/f)}{\sqrt{(1/f)^2+\tau^2}},  \quad \quad   \tan \delta \propto (1/f)^{-1}.\label{equ90}
\end{equation}
At low frequencies and $T > 0$, an power law may describe the complex susceptibility in the {\it creep} state. If one assumes $\chi=\chi_{\infty}[1+(if\tau)^{-\beta}]$ with $0 < \beta < 1$,
\begin{equation}
   \begin{array}{ll}
     \chi' - \chi_{\infty} \propto (1/f\tau)^{\beta}\cos (\beta \pi/2)   & \quad  \\
    \chi'' \propto (1/f\tau)^{\beta}\sin (\beta \pi/2) & \quad
   \end{array},
\label{equ100}
\end{equation}
which is depicted by a straight line $\chi''=(\chi'-\chi_{\infty})\tan(\beta \pi/ 2)$ in the Cole-Cole plot \cite{bra05,kle07}. In the low-frequency regime, $\chi (f,T=0)\approx \chi_{\infty}$. Thus a power-law behavior of $D\chi$ 
is expected from Eq.~(\ref{equ100}). The intermediate regime between the relaxation and creep states looks complicated \cite{pet04,fed04,bra05,kle07,kle06b}.

\begin{figure}
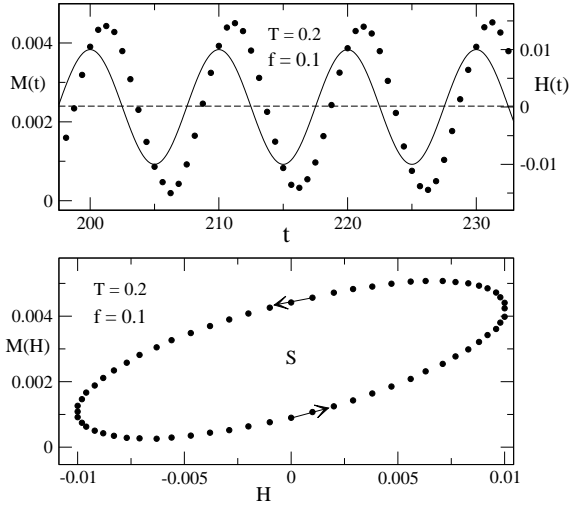
\onefigure[scale=0.4]{m_t.eps}
\caption{The upper panel shows the time oscillation of the magnetization (circles), in comparison to the external field (the solid line) with $H_0 = 0.01$, $f = 0.1$ and at $T = 0.2$. The lower panel demonstrates how the 
magnetization is driven by the external field. The area of the loop, $S$, is equal to $\chi''$.} \label{f0}
\end{figure}

In Fig.~\ref{f0}, the time oscillation of the magnetization is displayed for $H_0 = 0.01$, $f = 0.1$ and at $T = 0.2$ in the upper panel. Compared with the external field $H(t) = H_0 \exp(i2\pi ft)$, there exists a phase shift 
$\delta$ as indicated in Eq.~(\ref{equ60}). Both the amplitude and phase shift of the magnetization are generally frequency-dependent, and could be also temperature-dependent. In the lower panel, it is illustrated how the 
magnetization is driven by the external field. One can show that the area of the loop, $S$, is equal to $\chi''$.

\begin{figure}
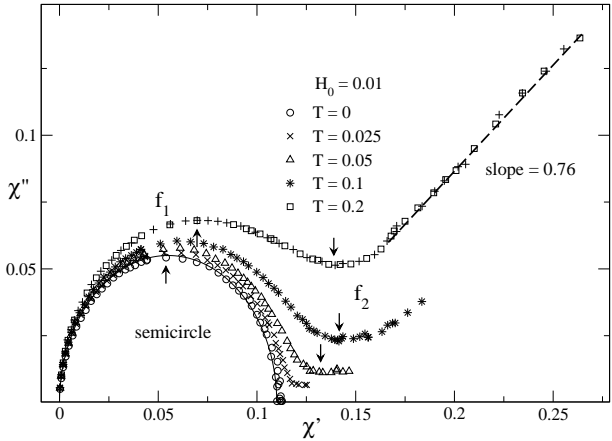
\onefigure[scale=0.35]{x1_x2.eps} \caption{Cole-Cole plot of the complex susceptibility for different temperatures and frequencies of the ac driving field. The solid line shows a semicircular fit, and the dashed line 
shows a linear fit. The characteristic frequencies$f_1$ and $f_2$, marked with up and down arrows, denote the local maximum and minimum of $\chi''$. Pluses show data for $L_x=24$ at $T= 0.2$.} \label{f1}\end{figure}

In Fig.~\ref{f1} we show the Cole-Cole plot of the complex susceptibility for the DRFIM model in a weak field $H_0 = 0.01$ at low temperatures. A perfect semicircle for the relaxation state is observed at $T=0$, consistent with 
Eq.~(\ref{equ80}) and recent experiments \cite{bra05,kle07,jez08}. As T increases, the creep motion gradually dominates and a linear function is observed at $T =0.2$ with a slope $0.76(3)$. The exponent $\beta = 0.41(1)$ is then 
calculated from $\tan(\beta \pi / 2) = 0.76(3)$. Two characteristic frequencies $f_1$ and $f_2$ can be identified by the local maximum and minimum of $\chi''$. In experiments, either $f_1$ or $f_2$ is considered as the transition 
point \cite{bra05,kle07}. In fact, these two frequencies separate three different regimes, the relaxation state $f > f_1$, creep state $f < f_2$, and intermediate state $f_2 < f < f_1$.

\begin{figure}
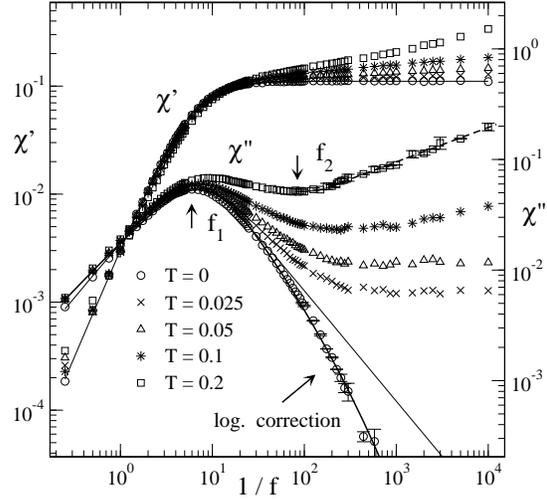
\onefigure[scale=0.4]{x_f.eps} \caption{The spectra of $\chi'$ and $\chi''$ on a log-log scale. Solid lines show the Debye law and its logarithmic correction for the relaxation state, and the dashed line indicates a 
power-law fit. Frequencies $f_1$ and $f_2$ are marked for $T = 0.2$.} \label{f2}\end{figure}

In Fig.~\ref{f2},  $\chi'$ and $\chi''$ are plotted as functions of $1/f$. Note that $f_2(T \to 0) = 0$ while $f_1( T \to 0) = 0.17$ Hz. At $T = 0$, the Debye law for the relaxation state in Eq.~(\ref{equ80}) is shown with solid 
lines. The relaxation time $\tau \approx 7$ is obtained from the fitting, equal to $1/f_1$. More accurately, a logarithmic correction is detected for $\chi''$, which has been reported as a non-Debye-type contribution \cite{fed04}. 
In the relaxation regime, the curves of different $T$ almost overlap, indicating that the relaxation motion is $T$-independent. In the creep regime, power-law behaviors of $\chi'$ and $\chi''$ in Eq.~(\ref{equ100}) are evident; 
however, measurements of the exponent $\beta$ are not precise, partly due to the existence of the relaxation motion at $T=0$. Additionally, the exponent $\beta$ estimated from $\chi''$ is smaller than that from $\chi'$ and in 
Fig.~\ref{f1}.

\begin{figure}
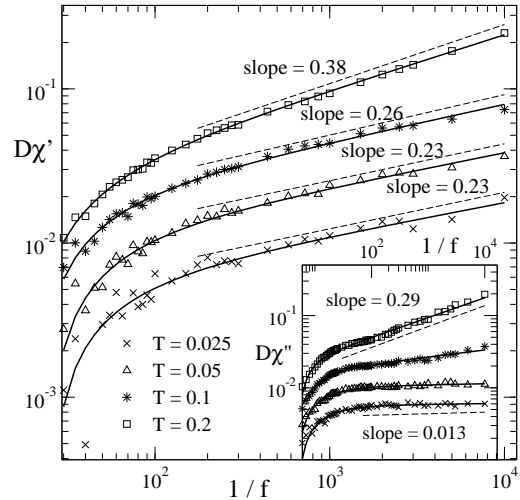
\onefigure[scale=0.4]{dx_f.eps} \caption{The spectra of $D\chi'$ on a log-log scale. Dashed lines represent power-law fits, and solid curves include corrections $y = a x^{\beta}(1 - c / x)$. The inset shows the 
spectra of $D\chi''$.} \label{f3}\end{figure}

To identify the pure creep motion induced by the temperature, we plot $D\chi'$ and $D\chi''$ as functions of $1/f$ at different temperatures in Fig.~\ref{f3}. Power-law behavior is observed, and the slopes of $D\chi'$ yield $\beta 
= 0.23(1)$, $0.23(1)$, $0.26(2)$ and $0.38(1)$ for $T = 0.025$, $0.05$, $0,1$ and $0.2$, respectively. $\beta=0.38(1)$ at $T = 0.2$ is consistent with $0.41(1)$ obtained in Fig.~\ref{f1}. A power-law correction may extend the 
fitting to the numerical data. Similarly, from $D\chi''$, we find $\beta=0.013(2)$, $0.027(4)$, $0.106(11)$ and $0.288(12)$, respectively. The difference of the exponent $\beta$ measured from $D\chi'$ and $D\chi''$ modifies the 
empirical ansatz in Eq.~(\ref{equ100}). $\beta (T\to0) \approx 0.2$ of $D\chi'$ and $\beta (T\to0) \approx 0$ of $D\chi''$ indicate that the creep motion induced by the temperature is different in $\chi'$ and $\chi''$. Finally, an 
approximate relation $D\chi(T)\propto T$ is detected in the creep regime when $T$ is low enough. \revision {According to Eq.~(\ref{equ100}), this is qualitatively consistent with the temperature dependence of the relaxation time 
reported in the literatures \cite{nat90,kle02}, $\tau \sim exp(c/T)$ }.

\begin{figure}
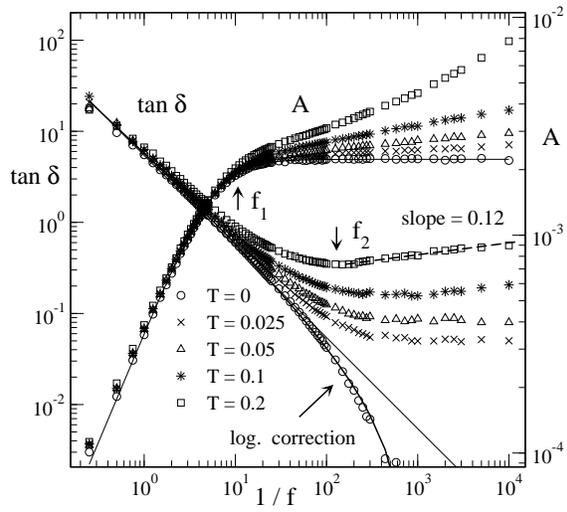
\onefigure[scale=0.4]{phase.eps} \caption{The amplitude $A$ and phase shift $\tan \delta$ for the magnetization as functions of $1/f$ on a log-log scale. The dashed line is a power-law fit, and the solid lines show 
the Debye law and its logarithmic correction in the relaxation state. $f_1$ and $f_2$ are the transition points for $A$ and $\tan \delta$ respectively. } \label{f4}\end{figure}

To capture the features of the intermediate state between the relaxation and creep states, we examine the dynamic behavior of the amplitude $A$ and the phase shift $\tan \delta$ of the magnetization, as shown in Fig.~\ref{f4}. At 
$T = 0$, Eq.~(\ref{equ90}) derived from the Debye law fits the data, and including a logarithmic correction improves the fitting of the phase shift $\tan \delta$. At $T=0.2$, the tail of $\tan \delta$ in the creep regime exhibits 
power-law behavior with a {\it nonzero} slope $0.12(1)$, which reflects the difference of the exponent $\beta$ for $\chi'$ and $\chi''$ obtained in Fig.~\ref{f3}. Interestingly, we observe that the frequency $f_1$ characterizes the 
transition behavior of $A$, while $f_2$ depicts that of $\tan \delta$, i.e. the relaxation-to-creep transition occurs in two stages, somewhat similar to the scenario of the two-dimensional melting \cite{lin06}, where an 
intermediate \revision {state} exists between the solid and liquid. In the relaxation regime, i.e., $f>f_1$, the domain wall \revision {relaxes microscopically without macroscopic movement}, the amplitude $A$ is $T$-independent, 
and the phase shift $\tan \delta$ decreases with $1/f$ although it is weakly $T$-dependent. For $f<f_1$, the creep motion induces the $T$-dependent amplitude $A$. Within the intermediate regime, i.e., $f_2<f<f_1$, however, the 
phase shift $\tan \delta$ decreases continuously with $1/f$, and the domain wall still looks partly like the relaxation state, \revision {especially at low temperatures}. At $f=f_2$, the phase shift $\tan \delta$ reaches its 
minimum. In the creep regime, i.e., $f<f_2$, the domain wall creeps macroscopically, the amplitude $A$ shows a linear behavior $A(T) \approx aT+b$ at low temperatures, and the phase shift $\tan \delta$ has a power law dependence on 
$1/f$.

In the literatures, the average magnetization $\overline M =(\omega/2\pi)\oint M(t)dt$ is often referred as the dynamic order parameter \cite{kle07,rob08}, and the area of the magnetization hysteresis loop $S = (\omega/2\pi) \oint 
M(H)dH$ is also concerned \cite{jez08}. By the definitions, $S \sim \chi''$ can be deduced, while $\overline M$ and $A$ show similar dynamic behaviors. From our numerical simulations, the amplitude $A$ and phase shift $\tan\delta$ 
may characterize the intermediate state more clearly.

\revision {In the creep state, it is well know that the relaxation time is induced by the metastable state of the domain wall, i.e., $\tau \sim exp(E_B/T)$ with $E_B$ being the energy barrier \cite{nat90}.
For the relaxation state, the Debye law in Eq.~(\ref{equ80}) can be derived with a simple two-level system \cite{nat90}. The only time scale $\tau$ in this case is still assumed to be the relaxation time of the metastable state. 
Such an assumption should be reasonable at not very low temperatures, and is supported by experiments \cite{bra05,mac08}. At zero or very low temperature, the scenario can be different, at least from the view of the DRFIM model. At 
zero temperature, for example, we detect a relatively small relaxation time as shown in Fig.~\ref{f2}. The domain wall relaxes and oscillates between the stationary states corresponding to the dc driving fields $H=\pm h_0$ and 
$H=0$, and the relaxation time is not induced by the metastable state with the energy barrier $E_B$. Qualitatively, it may be estimated by suddenly quenching the domain wall from the state of $H=0$ to $H=h_0$, and measuring the 
relaxation time of the domain-wall velocity. From our numerical simulations, we observe that the relaxation time $\tau$ increases with $h_0$, while decreases with the coupling constant $J$ and the disorder parameter $\Delta$. 
Therefore, the temperature dependence of $\tau$ varies in the parameter space of $h_0$, $J$ and $\Delta$. To well describe the delicate behavior in the experiments \cite{bra05,mac08}, however, more realistic interactions should be 
included.}

In summary, Monte Carlo simulations elucidate domain-wall motion in the two-dimensional random-field Ising model in an ac driving field. The relaxation and creep states are bounded by different characteristic frequencies $f_1$ and 
$f_2$ of the ac driving field: At $f = f_1$, the amplitude $A$ transits from $T$-independent to $T$-dependent, while at $f = f_2$, the phase shift $\tan \delta$ reaches its minimum. The complex susceptibility $\chi=\chi'-i\chi''$, 
the amplitude $A$ and phase shift $\tan\delta$ of the oscillating magnetization were examined at low temperatures. The Cole-Cole plot agrees characteristically with experiment, and power-law behaviors of $D\chi'(1/f)$ and 
$D\chi''(1/f)$ in the creep regime were observed. At $T=0.2$, the exponent $\beta = 0.38(1)$ obtained from $D\chi'$ is consistent with $\beta = 0.41(1)$ measured from $\tan(\beta \pi /2) = 0.76(3)$; however, $\beta = 0.29(1)$ 
estimated from $D\chi''$ is smaller and modifies the empirical ansatz in Eq.~(\ref{equ100}). Our values of the exponent $\beta$ can be compared with those from experiments,e.g., $\beta = 0.35(2)$ and $0.4$ in Refs. 
\cite{bra05,kle05}.

{\bf Acknowledgements:} This work was supported in part by NNSF of China grant Nos. 10875102 and 11075137, U.S. NSF grant No. DMR-0810223, and the JRG program of APCTP.


\end{document}